# $^3$He Lung Imaging in an Open Access, Very-Low-Field Human MRI System

R. W. Mair[1], M. I. Hrovat[2], S. Patz[3], M. S. Rosen[1], I. C. Ruset[4], G. P. Topulos[3], L. L. Tsai[1,5], J. P. Butler[6], F. W. Hersman[4], and R. L. Walsworth[1]

[1] Harvard-Smithsonian Center for Astrophysics, Cambridge, MA 02138, USA.

[2] Mirtech, Inc., Brockton, MA 02301, USA.

[3] Brigham And Women's Hospital, Boston, MA, 02115 USA.

[4] Department of Physics, University of New Hampshire, Durham, NH, 03824, USA.

[5] Harvard-MIT Division of Health Sciences and Technology, Cambridge, MA, 02139, USA.

[6] Harvard School of Public Health, Boston, MA, 02115, USA.

**Running Title:** Orientation-Dependent $^3$He Human Lung Imaging

**Corresponding Author:**

Ross Mair

Harvard Smithsonian Center for Astrophysics,

60 Garden St, MS 59,

Cambridge, MA, 02138,

USA

Phone:  1-617-495 7218

Fax:  1-617-496 7690

Email: rmair@cfa.harvard.edu

# ABSTRACT

The human lung and its functions are extremely sensitive to gravity, however the conventional high-field magnets used for most laser-polarized $^3$He MRI of the human lung restrict subjects to lying horizontally. Imaging of human lungs using inhaled laser-polarized $^3$He gas is demonstrated in an open-access very-low-magnetic-field (< 5 mT) MRI instrument. This prototype device employs a simple, low-cost electromagnet, with an open geometry that allows variation of the orientation of the imaging subject in a two-dimensional plane. As a demonstration, two-dimensional lung images were acquired with 4 mm in-plane resolution from a subject in two orientations: lying supine, and sitting in a vertical position with one arm raised. Experience with this prototype device will guide optimization of a second-generation very-low-field imager to enable studies of human pulmonary physiology as a function of subject orientation.

# INTRODUCTION

The lung is exquisitely sensitive to gravity, which may cause regional differences in blood flow, ventilation, gas exchange, alveolar size and other parameters [1-4]. Few methods exist that allow detailed studies of lung function under varying gravitational conditions – or subject orientations. Thus, pulmonary physiology could benefit greatly from the development of minimally-invasive methods to quantify lung function in subjects at variable orientations.

In recent years, MRI of inhaled, laser-polarized $^3$He gas [5] has become one of the most powerful methods for studying lung structure and function [6,7]. This technique is now used with conventional, high-field (~ 1 T) MRI instruments to make quantitative maps of human ventilation [8,9], to acquire $^3$He diffusion maps that yield lung passage airway size [10,11], and to monitor $^3$He relaxation as an indicator of alveolar gas-space $O_2$ concentration [12], with applications to basic pulmonary physiology and many lung diseases [13]. However, the large superconducting magnets used in conventional high-field MRI systems restrict human subjects to lie in a horizontal orientation. Although some initial studies have shown that posture changes affect the lung physiology in a way that can be clearly probed by $^3$He MRI [14,15,16], only minimal subject reorientation is possible inside conventional MRI scanners.

In addition, the great size, weight and technical restrictions of clinical MRI systems demand patients be brought to the scanner. For many critically ill patients, e.g., with lung diseases such as Acute Respiratory Distress Syndrome (ARDS), the requirement of being moved from the Intensive Care Unit is dangerous, time consuming, and expensive. Additionally, patients with implants, prostheses or claustrophobia are unable to undergo MRI scans in traditional scanners for physical or psychological reasons. Thus, the potential medical benefits of laser-polarized $^3$He

MRI are not realized for many of the most needy patients, and as such physicians could benefit from a compact, portable MRI system that could monitor critically ill patients in intensive care.

To enable orientation-dependent lung imaging, and possibly a portable, minimally-invasive lung MRI system, an open-access, very-low-field MRI system was developed, employing a four-coil electromagnet and exploiting the practicality of laser-polarized $^3$He MRI at magnetic fields < 10 mT [17,18]. $^3$He nuclear spin polarization of ~ 30% can be created by one of two laser-based optical pumping processes [5,19] prior to the MRI procedure; and then high-resolution gas space imaging can be performed without the need of a large applied magnetic field. Such high spin polarization gives $^3$He gas a magnetization density similar to that of water in ~ 1 T fields, despite the drastically lower spin density of the gas. Thus the signal-to-noise ratio (SNR) of laser-polarized noble gas MRI in animal or human lungs is only weakly dependent on the applied magnetic field [18], and very-low-field MRI becomes practical.

Very-low-field (~ 2 mT) imaging of small samples of laser-polarized $^3$He gas has previously been demonstrated [17,18], with obtainable SNR and image resolution comparable to high-field clinical scanners. In addition, it was shown that the effects of magnetic susceptibility-induced background gradients in the imaging sample are greatly suppressed for low-field MRI, resulting in much longer spin decoherence times for $^3$He in excised animal lungs [17]: $T_2^*$ ~ 100 ms at 2 mT, ~ 5 ms at 1.5 T. Recently, other groups have recognized the benefits of low-field MRI of laser-polarized gas, and have demonstrated human lung imaging at magnetic fields of 0.1 T [20-21], 15 mT [22], and even 3 mT [23,24]. However, all these demonstrations employed either conventional, horizontal bore MRI magnets [20-23], which restrict the subject to a single orientation, or a vertically-oriented electromagnet that does not allow variation of subject orientation [24].

In this paper, an overview of our prototype, open-access, very-low-field human MRI system is provided. With this system, the subject is unrestricted by the magnet and gradient coils in two dimensions, and only requires the placement of close-fitting but moderately flexible RF coils on the chest. As a demonstration, two-dimensional human lung images are presented from a subject in two orientations - lying horizontally and sitting vertically with one arm raised. Additionally, ongoing development of an upgraded very-low-field MRI instrument is discussed.

## EXPERIMENTAL

*Imager Design*: Details of the design and operation of our prototype (first generation) very-low-field human MRI system will be presented elsewhere. Here, a brief overview of the design, and how it permits open-access and variable-body-orientation lung imaging, is provided.

The first-generation imager operated with an applied static magnetic field, $B_0$, ranging from 0 - 70 G (0 – 7 mT). The $B_0$ field was created by two pairs of vertically-oriented coils that measured 2 and 0.8 meters in diameter (See Fig. 1). The 2 m coils were held on a heavy aluminum frame and spaced ~ 80 cm apart. The smaller coils were placed ~ 1 m apart and were mounted outside the large coils [25]. Current was supplied to each pair of coils separately from two stable, high current power supplies. The large coils were typically fed with ~ 30 – 40 Amps, while the smaller coils operated with ~ 20% of the current being sent to the larger coils. Crude shimming to optimize $B_0$ homogeneity was possible by adjusting the current in the small coils slightly with the large coils at a fixed current. This coil layout generated a field uniformity of ~ 1000 ppm over a diametrical spherical volume of ~ 30 cm.

Planar gradient panels were designed and constructed to provide the pulsed magnetic field gradients, thus eliminating another restrictive cylindrical geometry found in clinical MRI scanners. These planar gradient panels were mounted just inside the 2 m $B_0$ coils, and thus allowed variable-orientation lung imaging in an open-access gap of width ~ 75 cm. The planar gradient panels were made of 40 μm thick copper foil tape placed on 3 mm thick fiberglass sheeting. Three panels, for *x*, *y* and *z* gradients, were mounted together on the framing of each large $B_0$ coil. The winding pattern was obtained by calculating the resultant magnetic field gradient using the software package Biot-Savart [Ripplon Software, Burnaby, Canada], and then optimizing by iteration to maximize gradient strength and linearity throughout a spherical volume of ~ 60 cm diameter. The resulting winding pattern was qualitatively similar to those obtained with more sophisticated techniques [26]. The gradients were powered by Techron 8606 gradient amplifiers, operating at up to 140 A. At maximum current, the *z* gradient panel provided 0.38 G/cm, while the *x* and *y* panels provided 0.15 G/cm gradient strength.

RF-frequency control was provided by a commercial Surrey Medical Imaging Systems (SMIS) console, modified with a mix-down stage, to produce frequencies between 50 and 200 kHz. RF pulses from the SMIS were fed to a home-theater amplifier [Outlaw Audio, Durham, NH] that provided up to 250 W of RF power in conjunction with a homebuilt second-stage amplifier. The pre-amplifier was a Stanford Research Systems model 560. For simplicity, the system used separate coils for $B_1$ transmission and detection. We employed a Helmholtz pair of ~ 50 cm inner diameter (ID) for $B_1$ transmission, while the receive coil was a modified cosine-theta design of ~ 30 cm ID. Ensuring the two coils remained orthogonal was important to reduce direct pick-up noise – thus the two coils were bolted together to maintain orthogonality. Both coils could be tuned to different frequencies with external capacitative "resonance boxes". The coils had quality factors $Q$ ~ 50 - 100, implying operating bandwidths of ~ 1- 2 kHz at the typical $^3$He Larmor frequency ~ 100 kHz. Such narrow bandwidths, considerably less than typical imaging spectral widths of 10 – 20 kHz, required images to be post-processed to remove the convolved effect of the frequency response of the coil.

To improve SNR, the $B_0$, gradient and $B_1$ coils were housed inside an RF shielded room of steel plate on 1 inch particle board. The room was designed to attenuate RF interference in the range 10 kHz to 10 MHz by up to 100 dB. The imager hardware is photographed in Fig. 1.

*MRI Techniques*: For the demonstration human lung imaging presented below, the $B_1$ coils were tuned to 127 kHz, and $B_0$ set to 3.8 mT (38 G). Standard SMIS imaging sequences were modified for gradient echoes to minimize TE, and to permit non-sequential phase encoding and low flip angle excitation pulses. The RF pulses employed were nominally sinc-shaped and of 1 ms duration. For human imaging, the 2D (no slice selection) gradient echo sequence had a spectral width = 16.67 kHz, acquired field of view = 50 cm, maximum readout gradient of 0.1 G/cm, acquired data-set of 128 × 64 points (giving an in-plane resolution of ~ 4 × 8 mm), excitation flip angles of ~ 8º, TE/TR ~ 10/100 ms, and 1 signal averaging scan. The dataset was zero-filled to 256 × 256 points before fast-Fourier-transformation.

*Polarized $^3$He Production and Delivery*: Laser-polarized $^3$He gas for human lung imaging was produced via spin-exchange optical pumping using Rb vapor [5]. Our modular $^3$He polarization apparatus included gas storage, transport, and delivery stages that were similar to the $^{129}$Xe polarization instrument described in [27]. The polarization cells were ~ 80 cm$^3$ in volume, and made of GE-180 aluminosilicate glass. Each cell was enclosed in its own, dedicated Pyrex outer jacket that served as an oven, and attached to the polarizer (and hence the gas supply and storage) with a high-pressure Pyrex valve. The polarization cells could be easily removed from the polarizer, allowing diagnostic $^3$He NMR to be performed directly on the cell.

A magnetic field of ~ 1 mT was generated by a Helmholtz coil pair mounted on the polarizer, thereby providing a quantization axis for optical pumping. The polarizer was located adjacent to, but outside, the RF shielded room. For each experiment, we filled a polarization cell with ~ 5 - 6 bar of $^3$He and 0.1 bar of $N_2$, heated the cell to ~ 200°C, and applied ~ 60 W of circularly polarized light at 795 nm, provided by two fiber-coupled laser diode arrays [Coherent, Inc., Santa Clara, CA].

After spin-exchange optical pumping for ~ 2 - 4 hours, the $^3$He nuclear spin polarization reached ~ 20 – 40 %. The polarized gas was then expanded from the pumping cell into a previously evacuated glass and Teflon compressor for storage and delivery. For human imaging, polarized $^3$He was delivered via Teflon tubing through a feedthrough in the RF shielded room to a delivery manifold adjacent to the subject. This manifold consisted of a Tedlar bag, vacuum and inert gas ports, and a Teflon tube used as a mouthpiece. The $^3$He $T_1$ in both the compressor and the Tedlar bag was ~ 20 min.

*Human Imaging Protocol*: The subject in the demonstration human lung imaging was a healthy, 47 year-old male. Fig. 2 shows the subject in the very-low-field magnet, in both horizontal and vertical orientations. Just after a relaxed expiration, the subject inhaled, through the Teflon mouthpiece attached to the Tedlar bag, ~ 500 cm$^3$ of laser-polarized $^3$He gas, followed by a small breath of air to wash the helium out of the large airways and distribute it throughout the lung. The MR imaging sequence began immediately after inhalation, while the subject maintained a breath-hold for ~ 20 – 30 seconds. All human experiments were performed according to a protocol approved by the Institutional Review Board at the University of New Hampshire.

## RESULTS AND DISCUSSION

Fig. 3 shows the demonstration human $^3$He MRI lung images, acquired without slice selection at an applied field strength of 3.8 mT (38 G), with the subject in a supine orientation and a vertical orientation with one arm raised. Both images have a coronal orientation, with the lungs viewed in an anterioposterior direction, i.e., the subject's right lung is on the left side of the images. The horizontal image (Fig. 3a) shows well defined lung-lobes, with a characteristic concave shape at the bottom as the diaphragm pushes against the lungs. A region of lower intensity at the medial aspect of the left lung is consistent with the location of the heart. The gas distribution is very uniform throughout both lobes, as expected for a healthy subject in this orientation. The vertical image (Fig. 3b) shows the lungs to be considerably distended in comparison to the horizontal case, and indicates a less uniform gas distribution than when the subject was horizontal.

The image quality and SNR provided by the prototype very-low-field MRI system was limited by several technical imperfections: (i) mediocre $B_0$ homogeneity, measured to be ~ 1000 ppm in a 30 cm diameter volume at the magnet's center, resulting in $T_2^*$ ~ 5 – 10 ms for $^3$He inhaled in human lungs; (ii) excessive noise in the receive coil due to cross-talk from the $B_1$ transmit coil; and (iii) poor heat dissipation in the planar gradient panels, which required long delays (TR ~ 100 ms) between rows of k-space. As a consequence of these imperfections, sufficient SNR (~ 30) for a 2D image with in-plane resolution of 4 mm could only be achieved without slice selection and with relatively long image acquisition times (~ 7 s). Hence, the demonstration lung images of Fig. 3 do not show airway structural detail, and the vertical image suffers from artifacts due to subject motion. However, the observable detail is similar to the image presented in [23].

The images of Fig. 3 were obtained after post-processing to remove the convolved response function of the RF coil from the raw image data. Due to the high $Q$, and corresponding low coil bandwidth (~ 1 kHz), of the RF coils, raw images suffered significant attenuation away from the center frequency of the RF coils. The response function was obtained from rows of the image data-set that only contained noise. The entire data-set was then divided by this function, resulting in a flat noise floor across the entire image [28]. The images were obtained without using a variable flip angle to ensure reproducible transverse magnetization from each successive RF pulse [29], as the excitation flip angle was sufficiently low to ensure minimal variation in magnetization over the first 40 – 50 phase-encoding rows of the image, resulting in an artifact-free image.

A readout gradient of ~ 0.10 G/cm, which is an order of magnitude lower than values traditionally used at high field for human lung imaging with laser-polarized $^3$He gas, was used to acquire the images of Fig. 3. At large $B_0$, gradients of ~ 1 G/cm are employed to ensure that the pulsed gradient fields dominate the susceptibility-induced background gradient fields in the human lung, which scale with $B_0$. By operating at much lower $B_0$, there is no longer a need for large readout gradients, provided a correspondingly longer echo acquisition time is used so as not to compromise image resolution relative to that obtained at high field. Additionally, the maximum gradient across the sample, and deviations from linearity, remained low in comparison to $B_0$, avoiding concomitant field effects [30].

A significant benefit of operating at ~ 100 kHz was evident in our first demonstration lung images; namely operating well below the frequency range in which "sample noise" dominates human MRI [18]. Indeed, we found that insertion of the subject into and out of the RF coils had no loading effect whatsoever, thereby eliminating tuning/matching errors in addition to making sample noise insignificant. Conversely, environmental noise is important at low frequencies. The frequency spectrum ~ 100 kHz is crowded, and many electronic components of an MRI system, from computer monitors to the gradient amplifiers, emit noise in this range. RF shielding and line filters for conventional MRI instruments are optimized to eliminate noise in the 10 – 60 MHz range, not below that. In our first generation system significant reductions in pick-up noise from the gradient amplifiers were obtained by employing passive and active inductors on the gradient lines – however, there is room for improvement on the second-generation system currently under development.

## CONCLUSION

In vivo human lung imaging of inhaled laser-polarized $^3$He gas is demonstrated in an open-access, very-low-field MRI instrument operating at an applied magnetic field of 3.8 mT (38 G). A subject was imaged both while lying supine, and sitting vertically with one arm raised. Image quality and SNR was limited by technical imperfections in the prototype very-low-field MRI system, including (i) mediocre $B_0$ homogeneity, (ii) cross-talk noise between the transmit and receive $B_1$ coils, and (iii) poor heat dissipation in the pulsed field gradient panels. Nevertheless, the two-dimensional images show differences in lung shape, size, and gas distribution as a function of orientation. An optimized second-generation very-low-field instrument is currently being developed, which will enable detailed studies of lung function, ventilation and structure as a function of orientation. With further refinement, very-low-field MRI may be available in a portable, minimally-invasive instrument, allowing noble gas lung imaging for patients in intensive care units, as well as for those with implants, prostheses, claustrophobia or acute illnesses who have been denied access to MRI in its traditional form.

# ACKNOWLEDGEMENTS

Support is acknowledged from NASA grants NAG9-1166 and NAG9-1489, NIH grant RR14297, the Smithsonian Institution and University of New Hampshire. Equipment donations from Massachusetts Institute of Technology and Outlaw Audio, Durham, NH are gratefully acknowledged.

# FIGURE CAPTIONS

**Figure 1**. Photograph of the first-generation open-access, very-low-field human imager. The letters on the figure denote: A: the 2 m principal $B_0$ coils; B: the 80 cm secondary $B_0$ coils; C: planar pulsed-field gradient panels; D: $B_1$ transmit and receive coils for human imaging; E: support table for horizontal human imaging. The laboratory-frame Cartesian axes are also indicated. The design provides for a gap of ~ 75 cm for subject access between the planar gradient panels, and permits complete two-dimensional rotation of subjects in this plane, once the support table is removed. The entire apparatus is sited in a modular RF-shielded room.

**Figure 2**. Photographs of a subject in the first-generation open-access, very-low-field human imager. a) Subject on the support table, ready for imaging in the supine position. In this orientation, the two $B_1$ coils were bolted to the table, ensuring that their position remained independent of the subject. Only the ~ 50 cm diameter transmit coil is readily observable, the top of the receive coil can be seen under the subject's left hand. b) Subject sitting on a stool with one arm raised, ready for vertical orientation imaging. Removable supports hold the $B_1$ coils at the correct height, again ensuring their position was independent of the subject.

**Figure 3**. MRI of inhaled laser-polarized $^3$He gas acquired at $B_0 = 3.8$ mT (38 G) for the subject shown in Fig. 2. A standard FLASH gradient echo sequence was used, with TE/TR = 10/100ms, flip angle $\alpha = 8°$, NS = 1, data size = 128 × 64. Total image acquisition time was 7 seconds. a) Image acquired when the subject was lying supine. b) Image acquired while subject was sitting vertically with one arm raised. Both images visualize the lungs as if looking at the subject from the front – i.e., the subject's right lung lobe is on the left of the image.

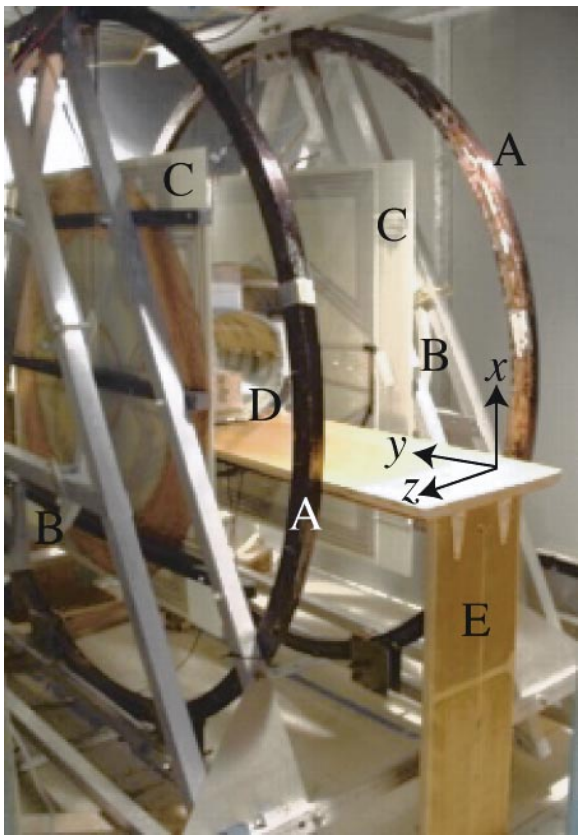

**Figure 1**

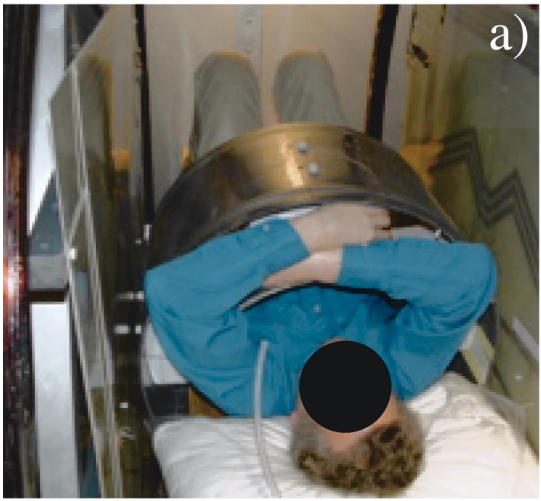 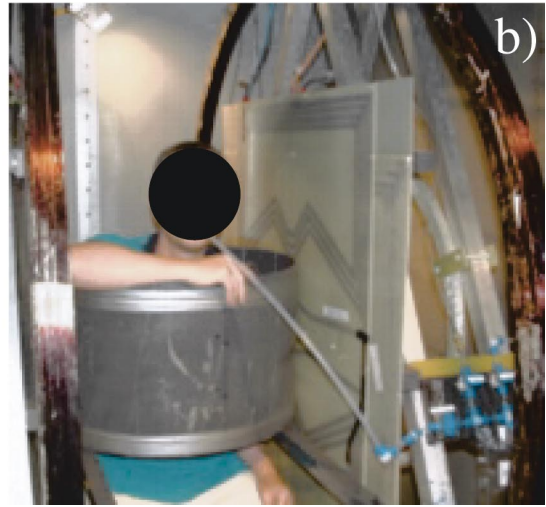

**Figure 2**

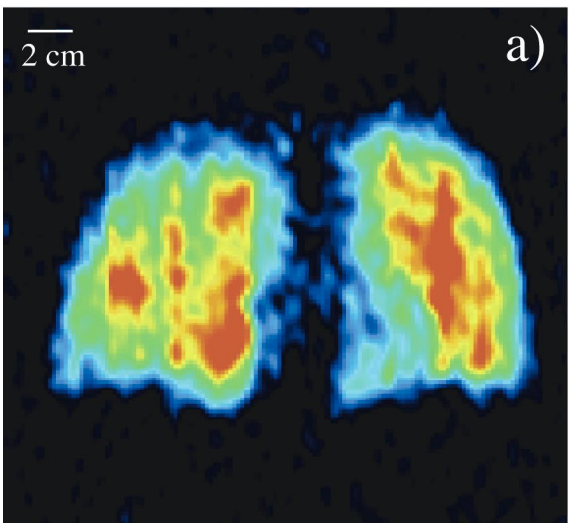 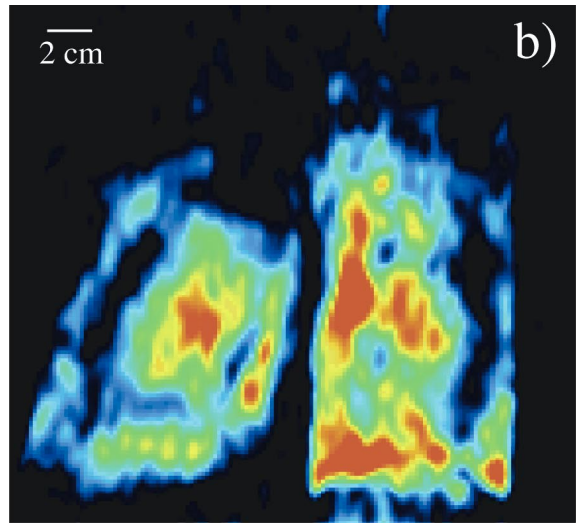

**Figure 3**